\documentclass[preprint,12pt]{elsarticle}

\usepackage{amssymb}

\journal{Physica A}

\begin{document}

\begin{frontmatter}


\title{Title\tnoteref{Fixing the fixed-point system}}

\title{Fixing the fixed-point system - Applying Dynamic Renormalization Group to systems with long-range interactions}


\author{Eytan Katzav \corref{cor1}}
\cortext[cor1]{corresponding author}
\address{Department of Mathematics, King's College London, Strand, London WC2R 2LS, United Kingdom}
\ead{eytan.katzav@kcl.ac.uk}
\begin{abstract}
In this paper a mode of using the Dynamic Renormalization
Group (DRG) method is suggested in order to cope with inconsistent
results obtained when applying it to a continuous family of
one-dimensional nonlocal models. The key observation is that the correct
fixed-point dynamical system has to be identified during the
analysis in order to account for all the relevant terms that are
generated under renormalization. This is well established for static
problems, however poorly implemented in dynamical ones. An
application of this approach to a nonlocal extension of the Kardar-Parisi-Zhang
equation resolves certain problems in one-dimension. Namely,
obviously problematic predictions are eliminated and the existing
exact analytic results are recovered.

\end{abstract}


\end{frontmatter}


\section{Introduction}
\label{sec:intro}
Fluctuating surfaces appear in a wide variety of physical situations
and have been of great interest in the last two decades
\cite{barabasi95,EW,KPZ}. These and other systems far from thermal
equilibrium pose a major challenge in contemporary statistical
physics. Behavior out-of-equilibrium is far richer than at
equilibrium, and many intriguing scaling phenomena, such as
self-organized criticality \cite{Bak}, or phase transitions between
non-equilibrium stationary states \cite{barabasi95}, have been
observed for long. However, despite the considerable achievements,
the theoretical comprehension of non-equilibrium phenomena remains
much poorer than our understanding of equilibrium phenomena.

The Renormalization Group (RG), proven useful to explain
universality in equilibrium continuous phase transitions, has also
allowed some progress in understanding systems out-of-equilibrium.
Nevertheless, in many cases the information RG analysis offers in
not complete and limited to a certain range of dimensions. A
classical example is the Kardar-Parisi-Zhang (KPZ) equation
\cite{KPZ} where the Dynamic Renormalization Group (DRG) approach
agrees with the analytic exact result in one dimension
\cite{barabasi95} but unable to provide results for the strong
coupling phase in higher dimension. This clearly indicates that
internal problems exist in the DRG calculation for $d>1$. Actually,
a remarkable result of Wiese \cite{Wiese98} shows that the
shortcoming of DRG in the KPZ system is not an artifact of a low
order calculation (so called "one loop" calculation), but rather
intrinsic to the method and extends to all orders. This situation
motivated the development of other methods to deal with the KPZ
system such as a scaling approach \cite{Hentschel91},
Self-Consistent Expansion (SCE) \cite{SE}, Mode-Coupling \cite{MC}
and others that were able to provide predictions for the exponents
in more than one-dimension.

A decade ago, a family of nonlocal growth models has been
introduced in \cite{Mukh97}, known as the Nonlocal KPZ (NKPZ)
equation, to account for nonlocal interactions in a system of
deposited colloids, giving rise to roughness larger than the one
predicted by the classical KPZ case. The authors studied the white
noise case that was later generalized to spatially correlated noise
in \cite{chat99}. To be more specific, the equation they studied was
\begin{equation}
\frac{{\partial h\left( {\vec r,t} \right)}}{{\partial t}} = \nu
\nabla ^2 h + \frac{\lambda_\rho}{2}\int {d^dr' \frac{\nabla h(\vec
r+\vec r') \cdot \nabla h(\vec r-\vec r')}{\left|2 \vec
r'\right|^{d-\rho}}} + \eta \left( {\vec r,t} \right)
 \label{eq:NKPZ} \, ,
\end{equation}
where $\eta \left(\vec r,t\right)$ is a noise-term modeling the
fluctuation of the rate of deposition, which has a zero mean and is
characterized by its second moment
\begin{equation}
\left\langle {\eta \left( {\vec r,t} \right)\eta \left( {\vec r',t}
\right)} \right\rangle  = 2D_0 \left| {\vec r - \vec r'}
\right|^{2\sigma  - d} \delta \left( {t - t'} \right)
 \label{eq:vareta} \, ,
\end{equation}
where $d$ is the substrate dimension and $D_0$ specifies the noise
amplitude. Note that in the limit $\rho \rightarrow 0$ the local KPZ
equation is recovered. This model suggests that the growth at each
point $\vec r$ gets contributions from pairs of gradients at points
symmetrically located around $\vec r$ along the interface, namely
$\nabla h(\vec r \pm \vec r')$, with a weight that is a decreasing function of the distance between them.

Both papers \cite{Mukh97,chat99} have investigated this problem
using the Dynamic Renormalization Group (DRG) approach, and have derived
a complex phase diagram. Focusing on the strong coupling solution
(in the KPZ sense \cite{barabasi95,KPZ}) both papers have found
\begin{equation}
z = 2 + \frac{{\left( {d - 2 - 2\rho } \right)\left( {d - 2 - 3\rho
} \right)}}{{\left( {3 + 2^{ - \rho } } \right)d - 6 - 9\rho }}
 \label{eq:zDRG} \, ,
\end{equation}
where $z$ is the dynamic exponent. The roughness exponent, $\alpha$,
characterizing the long distance spatial behavior, is obtained using
the modified Galilean scaling relation $\alpha + z = 2 - \rho$.
Unfortunately, the DRG result for the exponents, summarized in
Eq.~(\ref{eq:zDRG}) above, was found to be inconsistent with an
exact inequality in a certain range of the parameters and in all
dimensions \cite{inequality}. It is in place to comment here on the possible violations of the modified Galilean scaling relation, in view of recent criticisms of the relation between Galilean invariance and the scaling relation in the original KPZ system \cite{Berera,Nicoli,Wio10}. This may be especially relevant for discretized versions of Eq.~(\ref{eq:NKPZ}) (see Ref.~\cite{Wio10}) and less so in the continuum limit, which is the main focus of this paper.

Interestingly, another nonlocal extension of the KPZ equation has
been studied in the literature \cite{Nonlocal,NKPZ03}, namely
\begin{equation}
\frac{{\partial h\left( {\vec r,t} \right)}}{{\partial t}} = \nu \nabla ^2 h + \frac{\lambda_\rho}{2}\int {d^dr' \frac{\nabla h(\vec r') \cdot \nabla h(\vec r')}{\left|\vec r - \vec r'\right|^{d-\rho}}} + \eta \left( {\vec r,t} \right)
 \label{eq:NKPZ2} \, .
\end{equation}
This model also recovers the standard KPZ equation in the limit
$\rho \rightarrow 0$, but suggests that the growth at every point
$\vec r$ comes from the contribution of the gradients at all the points on the interface with a relative weight that decreases with the distance to $\vec r$. This is different from Eq.~(\ref{eq:NKPZ}) in that the nonlinearity contributes to growth via the local interaction with all the other points on the interface, and not just pairs of points
symmetrically distributed around it. 

It turns out that this model enjoys an exact result in $1D$ \cite{Nonlocal} predicting $z=(3-3\rho)/2$ when $\rho = 2\sigma$. It also happens that the same scaling relation $\alpha + z = 2 - \rho$ holds here, from which the roughness exponent $\alpha$ could be worked out. A more systematic study using the Self-Consistent Expansion \cite{NKPZ03} agrees with the exact result when applicable, and provides predictions for the exponents in other dimensions as well. On physical grounds this seems to be a simpler nonlocal extension of the KPZ nonlinearity than that of Eq.~(\ref{eq:NKPZ}), and therefore worthwhile understanding when modeling systems with long-range interactions.
This simplicity is reflected in the fact that the nonlinear term in Eq.~(\ref{eq:NKPZ2}) is more "relevant" (in the Renormalization Group sense) than the one in Eq.~(\ref{eq:NKPZ}), as will be seen below. 
The interesting thing is that the scaling dimension of the linear and non linear terms do not coincide in this equation, and this hinders the direct application of the perturbative Renormalization Group analysis.
A key observation made in the SCE analysis \cite{NKPZ03}, and which will be helpful for 
the DRG analysis as well, is that super diffusive modes of relaxation are generated by the nonlinearity of Eq.~(\ref{eq:NKPZ2}), namely super diffusion modes. This suggests that a remedy should be sought going back the old Renormalization idea of identifying first the right fixed point dynamical system around which the expansion should be. The fixed point dynamical system is not necessarily of the same form as the original system, as is implicitly assumed by the standard DRG procedure. That terms not included in the original action can be generated under the renormalization group has been known from the very beginning of the renormalization group, and was taken into account in static problems \cite{RG1,RG2,RG3}. However, this is often overlooked in dynamical problems.

In this paper a modification of the standard DRG procedure that goes
along those lines is suggested. This approach makes DRG more
flexible, and succeeds in recovering the exact result for the case
of the NKPZ Eq.~(\ref{eq:NKPZ2}). Not less important, this approach
could be useful in implementing DRG in other situations where
long-range interactions are present, such as those appearing in the
context of hydrodynamic interactions in colloidal suspensions
\cite{NMBE,NMBE2}, nonequilibrium fluctuations of an interface under shear \cite{Thiebaud},
wetting of an amorphous solid by a liquid
\cite{Golestanian,wetting} and in in-plane tensile crack propagation
in a disordered medium \cite{Ramanathan97,fracture}. The main
motivation here is to make the first step towards extending the range
of applicability of DRG in a field that suffers anyway from a lack
of analytical tools, in order to allow further progress in systems
out-of-equilibrium.

\section{Calculation and Results}
\label{sec:calc}
To understand the origin of the difficulty, consider the one loop
DRG. The renormalization procedure is most succinctly described
through the Fourier momentum $q$ and frequency $\omega$ modes, in
terms of which Eq.~(\ref{eq:NKPZ2}) becomes
\begin{equation}
h\left( {\vec k,\omega } \right) = G_0 \left( {\vec k,\omega }
\right)\eta \left( {\vec k,\omega } \right) + \lambda _\rho
{\mathcal{N}}\left[ {h\left( {\vec k,\omega } \right)} \right]
 \label{eq:RGiter1} \, ,
\end{equation}
where $G_0 \left( \vec k,\omega \right)$ is the bare propagator
given by $G_0 \left( \vec k,\omega \right) \equiv 1/(\nu_0 k^2 -
i\omega)$, and ${\mathcal{N}}\left[ {h\left( {\vec k,\omega }
\right)} \right]$ is a nonlinear functional of the height given by
\begin{eqnarray}
 && {\mathcal{N}}\left[ {h\left( {\vec k,\omega } \right)} \right] =  - \frac{1}{2}G_0 \left( {\vec k,\omega } \right)k^{-\rho} \\
 && \times \int {\int {\frac{{d^d \vec q d\Omega }}{{\left( {2\pi } \right)^{d + 1} }}\vec q \cdot \left( {\vec k - \vec q} \right)h\left( {\vec q,\Omega } \right)h\left( {\vec k - \vec q,\omega  - \Omega } \right)} } \nonumber
 \label{eq:N[h]} \, ,
\end{eqnarray}

The one loop expression for the dressed propagator defined by
$G\left( \vec{k},\omega  \right)\equiv {h\left( \vec{k},\omega  \right)}/{\eta \left( \vec{k},\omega  \right)}$ is given by \cite{barabasi95,medina89}
\begin{eqnarray}
 && G\left( {\vec k,\omega } \right) = G_0 \left( {\vec k,\omega } \right) + 4\left( { - \frac{\lambda }{2}} \right)^2 G_0 ^2 \left( {\vec k,\omega } \right) \\
 && \times \int {\int {\frac{{d^d \vec qd\Omega }}{{\left( {2\pi } \right)^{d + 1} }}2D_0 q^{ - 2\sigma } \left| {\vec k - \vec q} \right|^{-\rho} \left[ {\vec q \cdot \left( {\vec k - \vec q} \right)} \right]k^{ - \rho } } }  \nonumber \\
 && \times \left[ {\left( { - \vec q} \right) \cdot \vec k} \right] G_0 \left( {\vec k - \vec q,\omega  - \Omega } \right)G_0 \left( {\vec q,\Omega } \right)G_0 \left( { - \vec q, - \Omega } \right) \nonumber
 \label{eq:RGG1} \, ,
\end{eqnarray}
which, after some algebra (see appendix B in Ref.~\cite{barabasi95}, for example) gives
\begin{eqnarray}
 G\left(\vec k,0\right) &=& G_0 \left( {\vec k,0} \right) + \frac{{\lambda ^2 D_0 }}{{\nu _0^2 }}G_0^2 \left( {\vec k,0} \right)k^{2 - \rho } K_d \nonumber \\
 &\times & \frac{{d + \rho  - 2 - 2\sigma }}{{4d}}\int_{}^\Lambda  {dqq^{d - 3 - 2\sigma  - \rho } }
 \label{eq:RGG2} \, ,
\end{eqnarray}
where $K_d \equiv S_d/(2\pi)^d$, and $S_d$ is the surface area of a
$d$-dimensional unit sphere.

In the local KPZ case the last equation is used to calculate the
renormalization of the surface tension $\nu_0$. However, a look at
equation~(\ref{eq:RGG2}) highlights the problem. While the first
term on the RHS scales as $k^{-2}$ the second scales as
$k^{-2-\rho}$. Thus, a distinction between three cases should be
made: (a) When $\rho < 0$ the correction term (proportional to
$\lambda ^2$) is irrelevant compared to the first term in the limit
of small momentum (i.e. in the limit of large scales). (b) When
$\rho = 0$ both terms have the same scaling dimension. This
situation is actually the case in the classical KPZ equation (with
correlated noise), which is well studied for example in
Refs.~\cite{medina89,Katzav99}. And (c) when $\rho > 0$ the
correction is dominant over the first term. This means that in this
situation the perturbative expansion produces more relevant terms
than those originally present in the equation. More specifically a
fractional Laplacian is produced under the renormalization. This
implies that the fixed-point system in the space of dynamical
systems is a-priori not the original model, and one needs to
consider a more general form which contains such a term in the
equation in the first place. Adding a $\nu_1 k^{2 - \rho}$ term by
hand and going through the same process, a new (partially) dressed
propagator $G_1 \left(\vec k,\omega\right) = 1/\left( {\nu _1 k^{2 -
\rho } + i\omega }\right)$ is obtained. Repeating the steps
described above gives a $2^{nd}$ order expansion for the full
propagator, similar to Eq.~(\ref{eq:RGG2})
\begin{eqnarray}
 G\left( {\vec k,0} \right) &=& G_1 \left( {\vec k,0} \right) + \frac{{\lambda ^2 D_0 }}{{\nu _1^2 }}G_1^2 \left( {\vec k,0} \right)k^{2 - \rho } K_d \nonumber \\
 &\times& \frac{{d - 2\sigma  - 2 + 2\rho }}{{4d}}\int_{}^\Lambda  {dqq^{d - 3 - 2\sigma  + \rho } }
 \label{eq:RGG3} \, ,
\end{eqnarray}
This time, all the terms have the same scaling dimension so that the
perturbative expansion is meaningful in the sense that higher order
corrections are not more relevant than lower order ones. This allows
to calculate the renormalization of the effective surface tension
$\tilde \nu_1$ when $\rho > 0$.
\begin{equation}
\tilde \nu _1  = \nu _1 \left[ {1 - \frac{{\lambda ^2 D_0 }}{{\nu
_1^3 }}\frac{{d - 2\sigma  - 2 + 2\rho }}{{4d}}K_d \int_{}^\Lambda
{dqq^{d - 3 - 2\sigma  + \rho } } } \right]
 \label{eq:RGnu1} \, .
\end{equation}

Next, the renormalization of the noise term is calculated. The
effective noise $\tilde D$ is defined as the contraction of two
terms according to
\begin{eqnarray}
 &&\left\langle {h\left( {\vec k,\omega } \right)h\left( {\vec k',\omega '} \right)} \right\rangle = \\
 && = 2\tilde DG\left( {\vec k,\omega } \right)G\left( {\vec k',\omega '} \right) \delta^d \left( {\vec k + \vec k'} \right)\delta \left( {\omega  + \omega '} \right) \nonumber
 \label{eq:RGD1} \, .
\end{eqnarray}
The one loop expansion now yields
\begin{eqnarray}
 \tilde Dk^{-2\sigma} &=& D_0 k^{ - 2\sigma }  + \left( {2D_0 } \right)^2 \left( { - \frac{\lambda }{2}} \right)^2 k^{ - 2\rho } \nonumber \\
 &\times& \int {\int {\frac{{d^d \vec qd\Omega }}{{\left( {2\pi } \right)^{d + 1} }}} } \left[ {\vec q \cdot \left( {\vec k - \vec q} \right)} \right]^2 q^{ - 2\sigma } \left| {\vec k - \vec q} \right|^{ - 2\sigma } \nonumber \\
 &\times&\left| {G_1 \left( {\vec k - \vec q,\omega  - \Omega } \right)} \right|^2 \left| {G_1 \left( {\vec q,\Omega } \right)} \right|^2
 \label{eq:RGD2} \, .
\end{eqnarray}
Notice that $G_1 \left( {\vec k,\omega } \right)$ was used in the
last expression. In case $\rho < 0$ this should be $G_0 \left( {\vec
k,\omega } \right)$ as in standard DRG. Evaluating the integral in
Eq.~(\ref{eq:RGD2}) one obtains
\begin{equation}
\tilde Dk^{-2\sigma} = D_0 k^{-2\sigma} + k^{-2\rho} \frac{\lambda
^2 D_0^2}{\nu _1^3} \cdot \frac{{K_d }}{4}\int^\Lambda {dqq^{d - 3 -
4\sigma  + 3\rho } }
 \label{eq:RGD3} \,
\end{equation}
(or $\int^\Lambda {dqq^{d - 3 - 4\sigma}}$ when $\rho < 0$ ). As
before, the behavior of this equation is complicated by the
$k$-dependence of the correction term, and there are three options:
(a) When $\rho < \sigma $ the correction is irrelevant and the noise
amplitude does not renormalize. (b) For $\rho = \sigma$ the second
term is of the same order as the first term, and therefore
renormalizes the noise amplitude (the KPZ equation is an example of
this case since there $\rho = \sigma = 0$). And (c) for $\rho >
\sigma $ the correction term is more relevant than the first term.
This means that in this situation the perturbative expansion
produces an additional correlated noise which is more relevant than
that originally present in the equation. This implies, just like for
the propagator above, that the fixed-point system in the space of
dynamical systems does not have the form of the original model, and
a more general form with a new noise term $D_1 k^{-2\rho}$ is
considered. Doing the RG calculation from the beginning gives
\begin{equation}
 \tilde D = D_1 \left[ {1 + \frac{{\lambda ^2 D_1 }}{{\nu _1^3 }} \frac{K_d}{4}\int_{}^\Lambda  {dqq^{d - 3 - \rho } } } \right]
 \label{eq:RGD4} \, .
\end{equation}

Last, the one-loop contribution to the vertex $\lambda_\rho$ is
calculated. Without getting into all the details, the final results
is that the vertex does not renormalize to one-loop order $\tilde
\lambda_\rho = \lambda _\rho$, since the structure of the
perturbation theory is analytical in nature and cannot generate
singular terms that renormalizes $\lambda_\rho$. This is simpler
than the non renormalization of the vertex in the classical KPZ case
(with $\rho = 0$), where the correction is identically zero because
of some exact cancelation of terms (see Fig. B.3(c) in
\cite{barabasi95}).

Following the standard rescaling procedure the following flow
equations are obtained,
\begin{eqnarray}
 \frac{{d\nu _0 }}{{d\ell }} &=& \nu _0 \left( {z - 2} \right)\quad \quad \quad \quad \quad \quad \rho  < 0 \label{eq:RGnu2} \\
 \frac{{d\nu _1 }}{{d\ell }} &=& \nu _1 \left( {z - 2 + \rho  - K_d \frac{{\lambda ^2 D_0 }}{{\nu _1^3 }}\frac{{d - 2\sigma  - 2 + 2\rho }}{{4d}}} \right)\quad \rho  \ge 0 \nonumber
 \, ,
\end{eqnarray}
\begin{eqnarray}
 \frac{{dD_0 }}{{d\ell }} &=& D_0 \left( {z - 2\alpha  - d + 2\sigma } \right)\quad \quad \quad \quad \rho  < \sigma  \label{eq:RGD5} \\
 \frac{{dD_1 }}{{d\ell }} &=& D_1 \left( {z - 2\alpha  - d + 2\rho  + \frac{{K_d }}{4}\frac{{\lambda ^2 D_1 }}{{\nu _1^3 }}} \right)\quad \rho  \ge \sigma \nonumber
 \, ,
\end{eqnarray}
and
\begin{equation}
\frac{d\lambda_\rho}{d\ell} = \lambda_\rho (\alpha + z - 2 + \rho)
 \label{eq:RGlambda} \, .
\end{equation}

\begin{figure}[ht]
\centerline{\includegraphics[width=4cm]{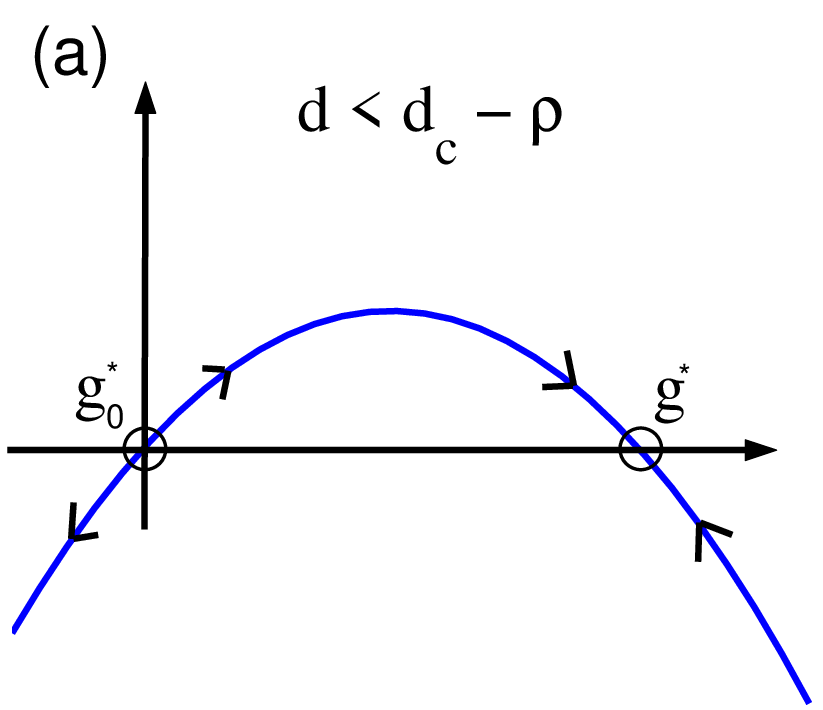} \includegraphics[width=4cm]{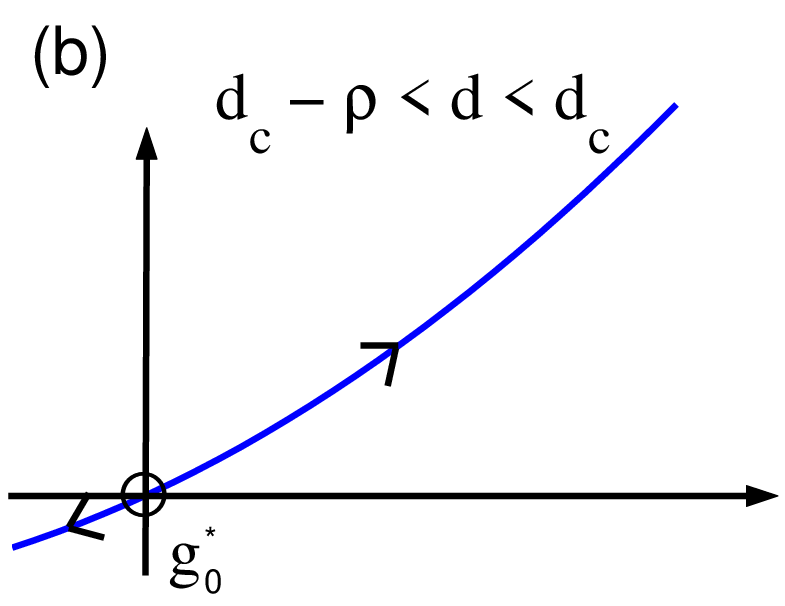} \includegraphics[width=4cm]{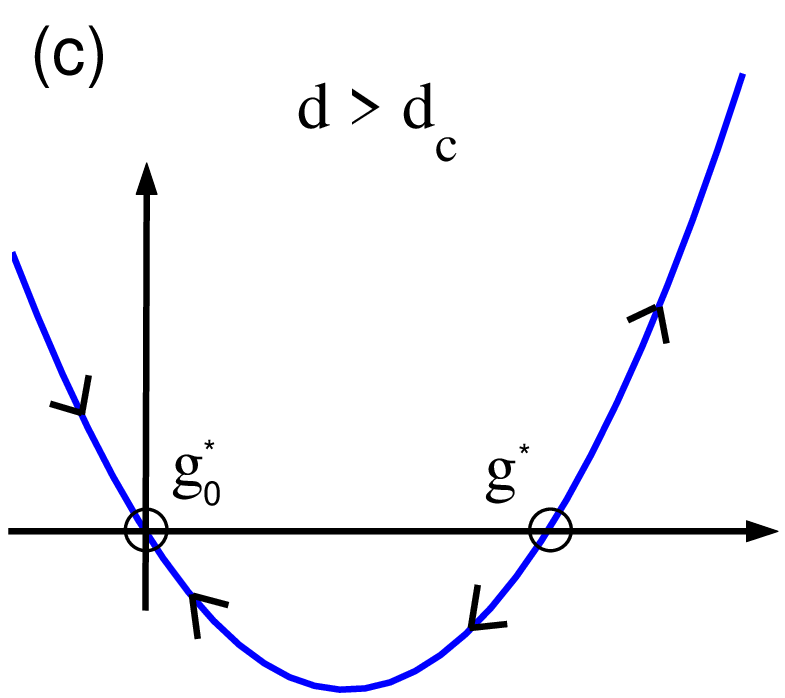}}
 \caption{Coupling constant flow for the NKPZ equation in Sector I ($\rho \ge 0$ and $\rho < \sigma$), where $d_c =2-\rho+2\sigma$. The three cases (a)-(c) cover all possible dimensions.}
 \label{fig:RGflow}
\end{figure}

The last step is a discussion of the complete RG flow for the NKPZ
equation. Four sectors in the $\rho,\sigma$-plane, in which
solutions can be looked for, are identified. In the following, a
detailed analysis in one of the sectors is presented, and results
for the other sectors are provided. Sector I is defined by $\rho \ge
0$ and $\rho < \sigma$. In this sector the flow equations are
(\ref{eq:RGnu2})b, (\ref{eq:RGD5})a and (\ref{eq:RGlambda}). As
traditionally done (in Refs.~\cite{barabasi95,medina89} for
example), it is simpler to combine the flow equation into one
equation for the coupling constant defined here as $g \equiv K_d
\lambda_\rho^2 D_0/\nu_1^3 d$. The RG flow of $g$ becomes
\begin{equation}
\frac{{dg}}{{d\ell }} = \left( {2 - d - \rho  + 2\sigma } \right)g +
3\left( {d - 2\sigma  - 2 + 2\rho } \right)g^2
 \label{eq:RGg} \, ,
\end{equation}
and so the Fixed Points (FP) for $g$ are
\begin{equation}
g_0^*  = 0 \quad \quad and\quad \quad g^*  = \frac{{2 - d - \rho  +
2\sigma }}{{3\left( {2 - d - 2\rho  + 2\sigma } \right)}}
 \label{eq:FPI} \, .
\end{equation}
A special dimension comes out of the last expression, the so-called
critical dimension which is $d_c  = 2 - \rho  + 2\sigma$. In Fig.
\ref{fig:RGflow} the RG flow of the coupling constant $g$ for
various dimensions is presented.

(a) The case $d < d_c - \rho $: As can be seen in
Fig.~\ref{fig:RGflow}(a), the nontrivial FP is the only stable FP in
this region, and by plugging it into the flow equations gives the
following scaling exponents $\alpha = (2 - d - \rho  + 2\sigma)/3$
and $z = (d + 4 - 2\rho - 2\sigma)/3$. These exponents are the
generalizations of the exponents of the classical KPZ system with
spatially correlated noise \cite{medina89,Katzav99}.

(b) For $d_c - \rho < d < d_c$ (depicted in Fig.~\ref{fig:RGflow}(b)), the
trivial FP $g_0^*$ is the only possible FP in the physical range as
the nontrivial FP is negative. However, $g_0^*$ is unstable, and so
the system flows towards $g = \infty$. However, just like the strong
coupling regime in the local KPZ equation, it is inaccessible to a perturbative
consideration, and one can just indicate its existence without
having a quantitative prediction for its scaling exponents.

(c) Last, as can be seen in Fig.~\ref{fig:RGflow}(c), for $d > d_c$
the trivial fixed point $g_0^* = 0$ is stable, and the exponents are
given by $\alpha  = (2 - d - \rho  + 2\sigma)/2$ and $z = 2 - \rho$.
These exponents correspond to the exponents of the Fractal
Edwards-Wilkinson equation (i.e. a linear equation with a fractional
Laplacian) with correlated noise. In addition, for a higher bare
value of the coupling constant $g > g^*$ , the system flows to $g =
\infty$, which signals the appearance of the strong-coupling regime,
again inaccessible to a perturbative approach. Thus, in this range
of dimensions there is a possible phase transition between a
weak-coupling to a strong-coupling regime.

The analysis of the possible phases in the other three sectors
follows the same lines and the results are summarized in
Table~\ref{tab:phases}. Note that for the strong coupling phases, no
quantitative result for the corresponding exponents is possible,
apart from pointing out their existence. Actually, we cannot even say 
if all the strong coupling regimes described here share the same scaling exponents.

\begin{table}
\begin{tabular}{@{}*{3}{l}}
 $\alpha$&$z$&validity\cr
 \hline
 $\frac{2-d-\rho+2\sigma}{2}$       & $2-\rho$ (FCEW)                          & $\rho \ge 0$, $\rho<\sigma$, $d>2-\rho+2\sigma$     \cr
 $\frac{2-d-\rho+2\sigma}{3}$       & $\frac{d+4-2\rho-2\sigma}{3}$            & $\rho \ge 0$, $\rho<\sigma$, $d<2-2\rho+2\sigma$    \cr
 Strong Coupling                    &                                          & $0 \le \rho \le \sigma$, $d>2-2\rho+2\sigma$    \cr
 $\frac{(2-d)(2-d+\rho)}{2(3-2d)}$  & $2-\rho-\frac{(2-d)(2-d+\rho)}{2(3-2d)}$ & $\rho \ge 0$, $\rho \ge \sigma$, $d<3/2$            \cr
 Strong Coupling                    &                                          & $\rho \ge 0$, $\rho \ge \sigma$, $d>3/2$            \cr
 $\frac{2-d-\rho+2\sigma}{2}$       & $2-\rho$ (FCEW)                          & $\rho \ge 0$, $\rho \ge \sigma$, $d>2+\rho$         \cr
 $\frac{2-d+2\sigma}{2}$            & $2$ (CEW)                                & $\rho <   0$, $\rho <   \sigma$                     \cr
 $\frac{2-d}{2}$                    & $2$ (EW)                                 & $\rho <   0$, $\rho \ge \sigma$, $d>2$              \cr
 Strong Coupling                    &                                          & $\sigma \le \rho <   0$                     \cr
\end{tabular}
\caption{A complete description of all the possible phases of the
NKPZ problem using the modified DRG scheme for any value of $d,\rho$
and $\sigma$. The first two columns give the scaling exponents
$\alpha $ and $z$ for a particular phase, and the third column
states each phase's validity condition. The values of the scaling
exponents in most of the strong coupling phases are not accessible using
DRG, and one can only indicate their existence. It is not even known
whether they correspond to the same phase, and thus described by the
same exponents or not.}
 \label{tab:phases}
\end{table}

As can be appreciated, the full description of the results given in
Table \ref{tab:phases} is quite rich. In order to gain more insight
into this special attention is given to the interesting one
dimensional case with $\sigma \ge 0$, where the following dynamic
exponent is obtained:
\begin{equation}
z = \left\{ \begin{array}{l}
 2\quad \quad \quad \quad \quad \rho  < 0 \\
 {{\left( {3 - 3\rho } \right)} \mathord{\left/
 {\vphantom {{\left( {3 - 3\rho } \right)} 2}} \right.
 \kern-\nulldelimiterspace} 2}\quad \quad \quad \rho  \ge 0 \\
 \end{array} \right.
 \label{eq:z1} \, .
\end{equation}
Note that for $\rho \ge 0$ the exact one-dimensional result is
recovered \cite{Nonlocal}, as was suggested in \cite{NKPZ03}, and
thus a major problem with DRG described above is solved. However,
the problem in more than one dimension is not solved, in the sense
that explicit prediction for the strong coupling exponents are not
available there. The only thing that can be said is in what region
of phase diagram such a strong coupling solution is expected. This
leaves the challenge of describing the correct phase diagram of such
models open.

In the same vein, attempts to apply the ideas developed here to the original Nonlocal KPZ equation (\ref{eq:NKPZ}) in order to cure the contradiction of Eq.~(\ref{eq:zDRG}) with the exact inequality \cite{inequality} were unsuccessful so far. The self-consistent expansion is to date the only
approach that is both consistent with the exact results in $1D$ and
provides predictions for the strong coupling exponents in higher
dimensions, as well as with the response-correlation inequality \cite{inequality}. Hopefully, this work will challenge more researchers to
derive results for the strong coupling regime using other promising methods such as the
Functional RG \cite{Fedorenko}, the Nonperturbative RG approach \cite{Canet}
or exploiting the recently proposed variational formulation \cite{Wio}
for KPZ-like systems. It also leaves a challenge for applying DRG
to more complicated systems, such as those exhibiting memory effects \cite{Chattopadhyay}.
It may be the case the higher order calculations would offer a solution to the inaccessibility to the strong coupling regime as done in \cite{RG4}. However, it is expected that in the KPZ context \cite{Wiese98} higher order terms would not solve the problem.


\section{Summary and Discussion}
\label{sec:sum}
To summarize, in this paper a modification of the classical DRG
approach to systems out of equilibrium is presented in order to
resolve problems with the results derived using traditional DRG for
the Nonlocal KPZ equation (\ref{eq:NKPZ2}). This approach extends beyond the NKPZ
system, to any system out-of-equilibrium that produces under
renormalization relevant terms which are not present in the original
model. For the NKPZ system (\ref{eq:NKPZ2}) (as well as for the Nonlocal Molecular
Beam equation \cite{NMBE} and the Fractal KPZ equation \cite{FKPZ1,FKPZ2})
it is found that for certain values of the parameters a fractional
Laplacian is generated under renormalization (i.e., for $\rho> 0$),
or a correlated noise term (i.e., for $\rho>\sigma$). Thus, an
inclusion of these terms in the original model (or put differently,
by considering the right fixed-point dynamical system) leads to a
correct description, and resolves the above-mentioned inconsistency
with the exact $1D$ result when $\rho = 2\sigma$. The importance of
including all relevant terms in the original action is well known
for many years, and was successfully applied to static problems
\cite{RG1,RG2,RG3}. However, not much attention to this effect has been given for
dynamical problems experiencing long-range interaction, where it
becomes important.

An interesting application of these ideas could be the case of a
driven wetting line of a fluid on a rough surface
\cite{Golestanian,wetting} or the mathematically similar problem of
an in-plane tensile crack propagating in a disordered material
\cite{Ramanathan97,fracture} (a moving rather than a pinned
interface). In these problems, a long-range interaction term exists
at the nonlinear order, and it is therefore vulnerable to similar
difficulties. As suggested in Refs.~\cite{wetting,fracture}, it
could be that the Edwards-Wilkinson system \cite{EW} is the
relevant fixed-point system for the rough phase of these physical
models. More work in that direction is needed to clarify this issue.








\end{document}